\documentclass[amssymb,twocolumn,%preprint,
aps,floatfix,showpacs]{revtex4}
\usepackage{graphicx}

\begin{document}
\title{Population trapping in bound states during IR-assisted ultra-fast
photoionization of Ne$^+$}
\author{H.W. van der Hart and R. Morgan}
\affiliation{Centre for Theoretical Atomic, Molecular and Optical Physics,
School of Mathematics and Physics,
Queen's University Belfast, Belfast, BT7 1NN, United Kingdom}

\date{\today}

\begin{abstract} 
We have investigated photoionization of Ne$^+$ in the combined field of a short
infra-red laser pulse and a delayed
ultra-short pulse of the infra-red laser's 23\textsuperscript{rd} harmonic. We
observe an ionization yield compatible with a picture in which one electron gets excited into
Rydberg states by the harmonic laser field and is
subsequently removed by the infra-red laser field. Modulations are seen
in the ionization yield as a function of time delay. These modulations originate from
the trapping of population in low members of the Rydberg series with different states being populated
at different ranges of delay times. The calculations further demonstrate that
single-threshold calculations cannot reproduce the Ne$^+$ photoionization yields obtained in
multi-threshold
calculations.
\end{abstract}


\maketitle

%%%%	Introduction	%%%
\section{Introduction}

One of the key challenges in atomic physics is the development of techniques to monitor
the motion of electrons in atoms, ions and molecules in detail. The typical timescale for
electron motion is in the order of 100 attoseconds. The development of experimental techniques to
produce ultra-short light pulses has been a key step towards meeting this challenge
\cite{Pau01,Hen01},
with current technology capable of producing pulses lasting 67 attoseconds \cite{Sho}. 

In combination with ultra-short light pulses, the application of streaking techniques
has proven to
be particularly valuable in making the dynamics apparent \cite{Kie04}. Using the combination of both techniques, 
experiment has been able to investigate a range of ultra-fast processes, including measurements of
relative time delays in atomic photoionization \cite{Sch10} and dynamics of shake-up states during high-frequency
ionization of noble-gas atoms \cite{Uib07}. In the latter
experiment, the observed dynamics is associated with dynamics within the residual ion. The ultra-fast
XUV laser pulse photoionizes Ne and leaves Ne$^+$ in a range of excited shake-up states. The IR pulse
then causes further ionization of these shake-up states, leading to double ionization. By varying the
timing of the XUV pulse with respect to the IR pulse, changes in the double ionization yield are
observed. These variations in the measured double
ionization signal can then be associated with the interplay between the excited Ne$^+$ ion and
the IR light field. The same process has also been studied in Kr atoms with longer excitation pulses,
with an estimated duration of 24-28 fs \cite{Bry12}. The difference in XUV pulse length leads
to subtle differences in the ionization yield. Whereas the ionization yield curves for long pulses
show smooth variation as a function of time delay, in the case of ultra-fast double ionization,
modulations can be noticed on top of the double ionization yields.

Several theoretical approaches have been applied to investigate ultra-fast dynamics in
excitation-ionization processes, or electron dynamics in excited Rydberg series. In \cite{Uib07}, the
obtained ionization yields were analysed under the assumption that the IR field ejects the emitted
electrons through tunnel ionization.  Argenti and
Lindroth investigated how time delays between an XUV-pulse and an IR pulse could steer ionization of He
towards either the He$^+$ 2s or the 2p state, following excitation of the Rydberg series associated
with these thresholds \cite{Arg10}. Dimitrovski and Madsen \cite{Dim08} studied the ionization of
dynamics of H using few-cycle laser pulses. In this study modulations in the ionization signal were
ascribed to half-cycle ionization dynamics. Kazansky and Kabachnik \cite{Kaz08} employed a
single-active-electron model for the description of the double ionization of Ne$^+$, where the
effect of the XUV pulse was described in the sudden approximation. They also observed oscillations
in the ionization yields, ascribed to quantum beats originating from coherent contributions of
eigenstates of the Ne$^+$ ion.

In this report, we aim to investigate a similar process, IR-assisted photoionization of the Ne$^+$
ion, where the XUV laser pulse excites Rydberg states below the Ne$^+$ ionization threshold. Dynamics
in the Rydberg series is then studied by varying the time delay of the XUV pulse with
respect to the IR pulse. However, in order to model the dynamics in detail, we aim to have an
accurate description of the Ne$^+$ atomic structure, as the interplay between Rydberg series associated
with different thresholds may affect the ionization dynamics. Calculations of high-harmonic generation
in noble-gas ions have demonstrated a factor 4 difference in the individual-atom harmonic yield obtained
from an ion initially with magnetic quantum number $M=0$ and an ion with $M=1$ \cite{Bro13,Has14}. This
difference follows from the structure in the residual doubly charged ion. Three different thresholds lie
close together, $n$p$^4$ $^3$P$^e$, $^1$D$^e$ and $^1$S$^e$. In the case of $M=0$, emission of an electron
towards the lowest $^3$P$^e$ threshold is reduced as emission of an $m=0$ is not allowed. A second
effect of the small separation between the thresholds is that the associated Rydberg series overlap. Hence,
Rydberg states will have admixtures from Rydberg states associated with different ionization thresholds.  

One of the approaches that can be used for the study of dynamics in atomic noble-gas ions from first
principles is time-dependent R-matrix theory \cite{Lys09,Nik08,Lys11,Moo11}. In the study of noble-gas ions,
this theory was first applied
to the study of two-photon ionization of Ne$^+$ in short high-frequency pulses, where it was
demonstrated how the two-photon ionization rate can be deduced from ionization yields, when absorption
of a single photon is just insufficient to achieve ionization \cite{Ham10}.
The theory has since been applied primarily to investigate the effect of the interplay between
closely spaced residual-ion thresholds on harmonic generation for Ar$^+$ and Ne$^+$
\cite{Bro12,Bro13,Has14}. These studies demonstrated that the
accurate description of harmonic generation requires at least inclusion of the $np^4$ $^3$P$^e$ and
$^1$D$^e$ thresholds. Recently, progress has been made in the description of
ultra-fast double ionization processes using time-dependent R-matrix theory \cite{Har14}, but
at present these processes can only be described for two-electron systems.

In the present report, we apply the most recent implementation of time-dependent R-matrix theory,
R-matrix including time dependence \cite{Lys11,Moo11}, to study IR-assisted photoionization of Ne$^+$
using ultra-short light pulses.
This implementation of time-dependent R-matrix theory
is significantly more efficient for the exploitation of massively parallel
computing facilities. This is of great assistance when dealing with IR fields, as intense
long-wavelength fields require the inclusion of many angular momenta, and hence extensive basis sets.
For time-dependent R-matrix approaches, this is an important consideration. The description
of multiphoton processes using limited amounts of atomic structure is much more reliable when the light
field is described in the length form \cite{Hut10}. The length form, however, necessitates the
inclusion of many more angular momenta when IR fields need to be considered \cite{Cor96}. 

We will start the report with a description of the various computational techniques and basis sets used
in the calculations. We will then provide Ne$^+$ ionization yields as a function of time delay. We
then analyse the final-state wavefunction in detail, and investigate how the population
of individual final states is affected by the time delay. We finish with our conclusions.

\section{Computational method}

R-matrix theory including time dependence uses the standard R-matrix approach by
separating the system into two distinct regions, known as the inner region and the outer
region \cite{Bur11}. In the inner region, all electrons are close to the nucleus. The electron-electron
interactions are strong and all interactions between all electrons are therefore
taken into account. In the outer region, one electron has moved away from the nucleus. For this
electron, exchange interactions involving the residual-ion electrons can be neglected. This
outer electron thus moves in a potential combining the long-range potential associated with the
residual ion and the laser field potential.

In standard R-matrix theory the two distinct regions are connected through the so-called R-matrix
\cite{Bur11}, but in R-matrix theory including time dependence, the connection between the two
regions is made through the wavefunction directly \cite{Nik08,Lys11,Moo11}. In the inner region, the
wavefunction is described using a large configuration-interaction basis. In the outer region, the
wavefunction is described using a direct product of a channel function, a residual-ion state and
spin-angular functions of the continuum electron, and a grid representation of the radial wavefunction
of the outer electron. The inner-region wavefunction is provided to the outer region, by extending
the outer-region
grid into the inner region. The inner-region wavefunction is then evaluated on this grid.
This provides all the information needed by the outer region to propagate the wavefunction in the
outer region. To propagate the wavefunction in the inner region, time derivatives of the outer-region
wavefunction are determined at the inner-region boundary. Additional time propagators are then
determined to propagate these time derivatives into the inner region.

Time propagation in the RMT approach is performed using Arnoldi propagators \cite{Smy98}. In the
outer region, a single propagator suffices to propagate the wavefunction. In the inner region, separate
Arnoldi propagators are constructed for the initial wavefunction, and for each of the time derivatives
determined at the boundary \cite{Lys11,Moo11}. The number of propagators needed is given by the order
of the Arnoldi propagator. In the present calculations, we use Arnoldi propagators of order 12, and
a typical time step of 0.01 atomic unit. 
The grid spacing in the outer region is 0.08 $a_0$.
The spatial
operators are evaluated using five-point finite-difference rules.
The calculations
use a total of 864 processors, of which 784 are dedicated to the inner region and 80 are dedicated to
the outer region. The outer region covers a total distance of over 4900 $a_0$.

In the inner region, Ne$^+$ is described by a similar wavefunction expansion as used in a
previous comparison of R-matrix-Floquet theory and time-dependent R-matrix theory \cite{Ham10}. 
Following ionization of Ne$^{+}$, the residual Ne$^{2+}$ ion can be left in three states: 2s$^2$2p$^4$
$^3$P$^e$, $^1$D$^e$ and $^1$S$^e$. These states are generated using an orbital set containing
1s, 2s and 2p orbitals for the Hartree-Fock ground state of Ne$^{2+}$ \cite{Cle74}, and
additional correlation orbitals 3s, 3p, 3d, 4s, $\overline{\rm 4p}$ and $\overline{\rm 4d}$ orbitals
to improve the description of the residual ion states \cite{McL00}.
The Ne$^{2+}$ basis set contains the 1s$^2$2s$^2$2p$^4$,
1s$^2$2s2p$^5$ and 1s$^2$2p$^6$ configurations and all single excitations of a 2s and 2p electron from
these configurations. A small configuration-interaction calculation
then provides the desired Ne$^{2+}$ 2s$^2$2p$^4$ target states. The Ne$^+$ basis is
generated by combining these target states with an extensive set of continuum
orbitals.
These continuum orbitals are generated from a B-spline basis set of 50 B-splines of order 13.
The knot set spacing varies from nearly quadratic near the nucleus to nearly linear near the
outer-region boundary. To improve the description close to the nucleus, extra knot points are
inserted near the nucleus.
In addition, correlation orbitals are generated by combining the
Ne$^{2+}$ CI basis functions with an additional function from the orbital set. The maximum total
angular momentum retained in the calculation is $L_{\rm max} = 29$. The inner region boundary is
set at 15 $a_0$. The energy of the lowest field-free $^2$P$^o$ state is adjusted to match the
Ne$^+$ binding energy, and the energy of the lowest field-free $^2$S$^e$ state is also adjusted
to match the literature value \cite{Kra13}. The reason for this shift is that the basis
set for the Ne$^+$ states is larger compared to the one for the Ne$^{2+}$ states. Hence the lowest
Ne$^+$ states tend to lie too low with respect to the Ne$^{2+}$ thresholds. The Ne$^+$ ground state
is shifted up in energy by 0.17 eV, whereas the 2s2p$^6$ state is shifted up by 0.33 eV.

In the time propagation, we do not include all field-free eigenstates obtained in the inner-region
calculation. States with an energy exceeding 1345 atomic units above the first ionization threshold
are excluded from the calculation. These states have rapidly oscillating wavefunctions near the nucleus
and therefore are not physically relevant in the present study. However, at an energy of 1344 atomic
units above
the first ionization threshold, we obtain eigenstates localised at the inner-region boundary. These
states acquire a large eigenvalue arising from the Bloch operator, introduced in the inner-region
codes to make the inner-region Hamiltonian Hermitian \cite{Bur11}. These states describe aspects
of wavefunction flow through the boundary, and thus need to be retained in the calculation. The
RMT decides which states need to be included by monitoring the maximum boundary amplitude of each
eigenstate. Within a symmetry, all states up to and including the last state with a
sizable boundary amplitude are retained in the calculation, whereas all higher-lying states
are excluded. 

\begin{figure}
\includegraphics[width=7.9cm]{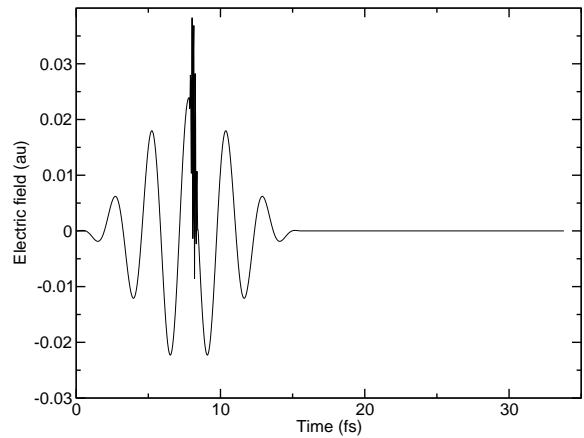}
\caption{Time profile of the combined electric field of the IR and the XUV pulse. The
shown delay of the XUV pulse is 7.8 fs, corresponding to 3 cycles of the IR field. Since
the delay is determined from the onset of the pulse, the delay means that the maximum in the
XUV pulse occurs when the IR field is decreasing.
}
\label{fig:pulse}
\end{figure}

The light field consists of two laser pulses, an IR field with a wavelength of 780 nm and an
XUV field involving the 23\textsuperscript{rd} harmonic of the IR field. Both fields last
for six cycles and have a $\sin^2$ profile. The oscillating electric field has a $\cos$-profile
so that the maximum of the envelope coincides with a maximum in the electric field.
After the IR pulse has finished, the wavefunction
is propagated for a time corresponding to 7 cycles of the IR field. A typical profile for the
electric field is shown in figure \ref{fig:pulse} for a time delay between the IR and the XUV
pulse of 7.8 fs (3 optical cycles). The delay time between the pulses is defined by the
onset of the $\sin^2$ profile. As a consequence figure \ref{fig:pulse} shows that, for
a delay of 7.8 fs (a 3-IR-cycle delay), the maximum of the XUV excitation pulse occurs about 0.34 fs
(0.13 IR-cycle) after the peak of the IR field.
The peak intensity for both pulses is set to $2\times10^{13}$ W/cm$^2$.

\section{Results and discussion}

In the present calculations, we aim to obtain ionization yields of Ne$^+$ in the combined field
of an XUV excitation pulse into the Rydberg series, and an IR pulse, which assists the excited
electron to fully escape. Since the initial excitation process involves excitation into Rydberg
states, it is important to account for the possibility that the residual Ne$^+$ ion has been left
in a highly excited state. In fact,
notable fluctuations are observed in the outer-region population due to wavefunction flow from the inner
region into the outer region and back. This is a clear sign that the population in Rydberg states
cannot be ignored. However, we also have to take into account that the Ne$^+$ ion
can be left in a Rydberg state associated with an excited threshold. This type of Rydberg state may
autoionize, leading indirectly to photoionization. We therefore approximate the
ionization yield by accounting for two different processes. When an electron is distanced further
than 115 $a_0$ from the nucleus, it is assumed to have escaped from the residual ion. When an
electron is at a distance between 15 and 115 $a_0$ in the outer region and in an outer-region
channel associated with an excited Ne$^{2+}$ threshold, it is assumed to be in an autoionising state,
resulting in delayed, or indirect, photoionization. In order to limit the dependence of the outcomes
on the final time used in the calculations, the ionization yields are averaged over the final
7.8 fs of the calculation (three cycles of the IR field).

\begin{figure}
\includegraphics[width=7.9cm]{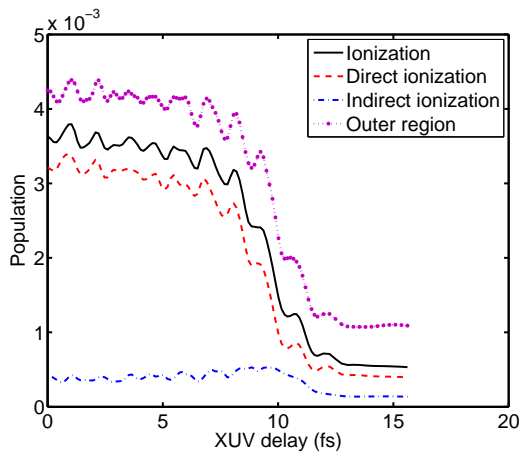}
\caption{(Colour online) Ne$^+$ ionization yield obtained as a function of time delay between the IR pulse and the XUV
pulse. The XUV pulse is the 23\textsuperscript{rd} harmonic of the IR pulse of 780 nm. The ionization
yield (solid black line) is given by the sum of the direct ionization yield (Direct ionization, red dashed
line) and the population left in Rydberg states associated with excited thresholds (Indirect ionization,
blue dot-dashed line). The difference between the ionization
yield and the total population in the outer region (purple dotted line with circles) can be assigned to
population in Rydberg states associated with the ground-state threshold.}
\label{fig:23ion}
\end{figure}

Figure \ref{fig:23ion} shows the yields, obtained as described above, associated with
photoionization and autoionization through excitation of a Rydberg state associated with an excited
Ne$^+$ threshold. For comparison, the total population in the outer region, averaged over the
same final-time interval, is also shown. It can be seen that about 80\% of the total population in
the outer region is associated with direct ionization. A substantial fraction of the outer-region
population therefore corresponds to excitation of Rydberg states. When the XUV pulse arrives prior
to the peak of the IR field, about 40\% of the Rydberg-state population corresponds to excited
thresholds, whereas if the XUV pulse arrives when the peak of the IR field has passed, only about 20\%
of the Rydberg-state population corresponds to excited thresholds. The figure demonstrates that a
significant fraction of the outer-region population is still confined to bound states, and that
these bound states can be attached to both the ground-state threshold and to excited thresholds.

Figure \ref{fig:23ion} shows a significant variation in ionization yield as the delay of the XUV
pulse is varied across the IR pulse. When the XUV pulse occurs at the tail end of the IR pulse, the total
probability for ionization amounts to about 0.6 $\times$ 10$^{-3}$, whereas the total probability
for ionization is a factor 6 larger when
the XUV pulse occurs at the start of the IR pulse. The largest decrease in the ionization yield
is seen at a delay of about 10 fs. Since this delay is well after the peak of the IR field at 7.8 fs,
the peak IR intensity of 2$\times$10$^{13}$ W/cm$^2$ is more than sufficient to strip electrons
from most of the Rydberg series.

One clear observation from figure \ref{fig:23ion} %, \ref{fig:thres} and \ref{fig:freq}
is that the ionization yields do not increase uniformly: for certain delay times local minima 
can be seen in the ionization yields. A similar behaviour was seen in the ultra-fast experiments
on photoionization with
excitation of Ne followed by IR photoionization of the excited Ne$^+$ ions. The origin of the 
oscillations has also been discussed  theoretically, through, for example, single-active-electron
calculations for model Ne$^+$ ions \cite{Kaz08}
and ultra-fast excitation-ionization studies of H \cite{Dim08}. We will now focus on the detailed
final-state wavefunction to investigate the role of different atomic states in the ionization dynamics.

\begin{table}
\caption{Energies of field-free states contained within the inner region with a
radius of 15 $a_0$. Energies from the present calculations are compared with literature
values \cite{Kra13}.}
\begin{ruledtabular}
\begin{tabular}{lll}
State & Excitation energy & Excitation energy \\
& Present & \cite{Kra13} \\
 & eV & eV\\ \hline
2s$^2$2p$^5$ $^2$P$^o$ & 0 & 0 \\
2s2p$^6$ $^2$S$^e$ & 26.877 & 26.878 \\
2s$^2$2p$^4$($^1$D)3s $^2$D$^e$ & 30.874 & 30.517 \\
2s$^2$2p$^4$($^3$P)3p $^2$P$^o$ & 31.597 & 31.485 \\
2s$^2$2p$^4$($^1$D)3p $^2$F$^o$ & 34.298 & 33.988 \\
2s$^2$2p$^4$($^1$D)3p $^2$P$^o$ & 34.534 & 34.231 \\
2s$^2$2p$^4$($^1$S)3s $^2$S$^e$ & 34.711 & 34.271 \\
2s$^2$2p$^4$($^3$P)3d $^2$D$^e$ & 34.800 & 34.727 \\
2s$^2$2p$^4$($^3$P)4p $^2$P$^o$ & 36.425 & 36.440 
\end{tabular}
\end{ruledtabular}
\label{tab:energy}
\end{table}

In order to analyse the origin of the modulations in the ionization yield, we take a more in-depth
look at the final-state wavefunctions, and the coefficients associated with the field-free
inner-region wavefunctions in particular. These populations are taken at the end of the calculation
with no averaging. The inner-region wavefunction is described in terms of
Ne$^+$ eigenfunctions, expressed in terms of R-matrix basis functions. If the state is relatively
tightly bound, the state is well contained within the inner region, and we can estimate the final
population in a particular state from its contribution to the final-state inner-region wavefunction. When the
state is less tightly bound, this approach will not be appropriate as a substantial fraction of
the wavefunction will extend into the outer region. For the present box size, we can use this
approach for 3s and 3p outer electrons, and for 3d and 4p electrons attached to the $^3$P$^e$
threshold. Energies and labels of the states investigated are given in table \ref{tab:energy},
and compared with reference data \cite{Kra13}. The differences in energy between the present
data and the literature values ranges up to about 0.5 eV. The main energy difference is associated
with energy differences in the residual-ion energies, with difference for states attached to the
$^3$P$e$ threshold around 0.1 eV, for states attached to the $^1$D$^e$ threshold around 0.35 eV, and
a difference of 0.44 eV for the state attached to the $^1$S$^e$ threshold. To improve on these
differences, even larger basis expansions for the description of Ne$^{2+}$ are required.

The final-state populations of the field-free Ne$^+$ ground state and the 2s2p$^6$ $^2$S$^e$ state
has also been investigated. For
these states, however, we observe only minor variations in the final-state population as a
function of XUV delay time, with fluctations of about 1.4\% in the final depopulation
of the ground state and fluctuations of about 3\% for the final population of the
2s2p$^6$ $^2$S$^e$ state. The lack of variation in the ground-state depopulation with respect to
XUV pulse delay also means that the variation in obtained ionization yield must be reflected
in a variation of population in excited states.

\begin{figure}
\includegraphics[width=7.9cm]{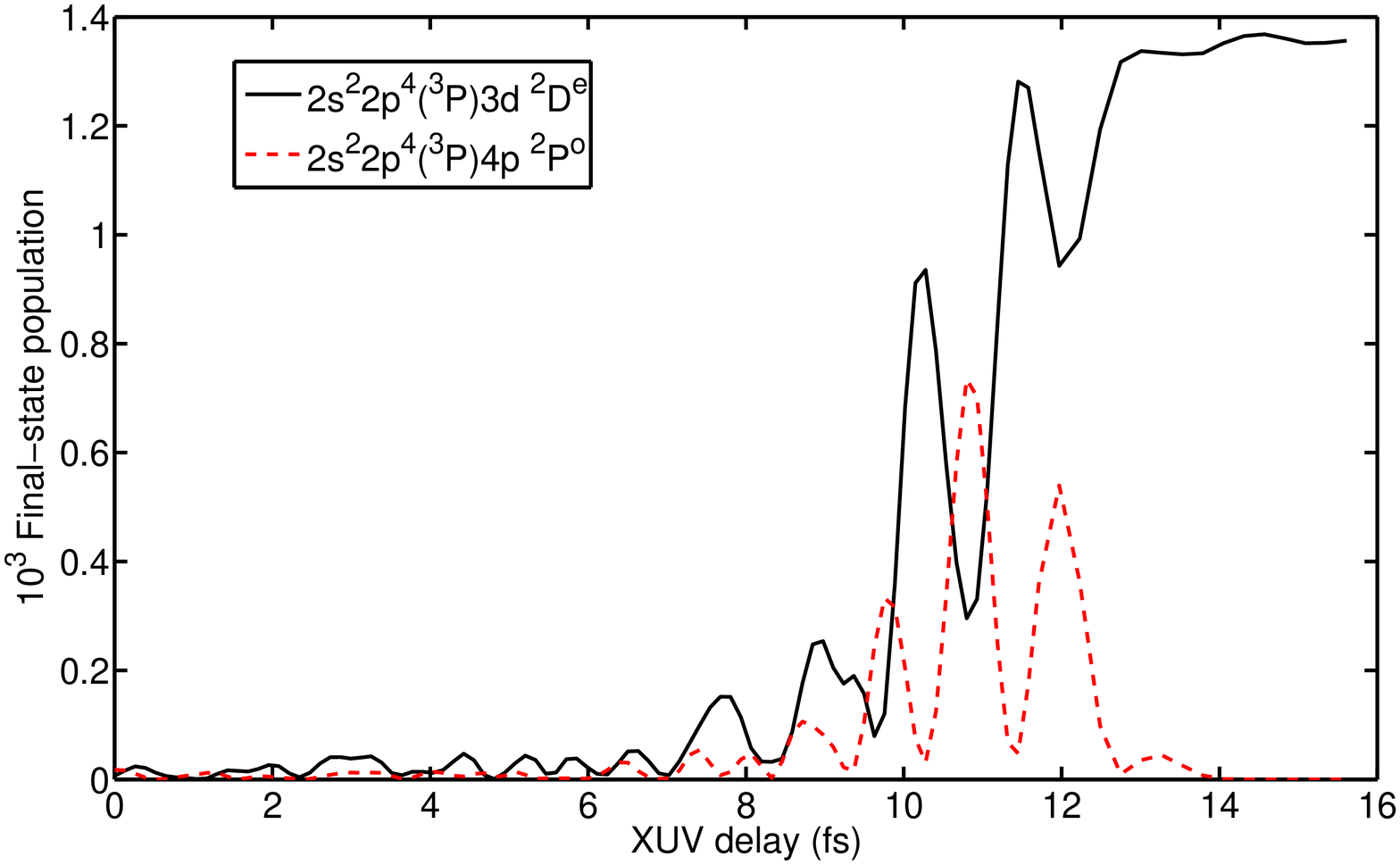}
\caption{(Colour online) Final-state population in 2s$^2$2p$^4$($^3$P$^e$)3d $^2$D$^e$ (black full line)
and 2s$^2$2p$^4$($^3$P$^e$)4p
$^2$P$^o$ (red dashed line) as a function of XUV time delay.}
\label{fig:3d}
\end{figure}

To illustrate how the IR field may remove population out of high-lying Rydberg states, we show
in figure \ref{fig:3d} the final-state population in 2s$^2$2p$^4$($^3$P$^e$)3d and
2s$^2$2p$^4$($^3$P$^e$)4p as a function of XUV time delay. The figure demonstrates that
2s$^2$2p$^4$($^3$P$^e$)3d receives about 25\% of the total population that is excited by the
XUV pulse. The figure demonstrates that when excitation occurs during the IR pulse, the final-state
population in 2s$^2$2p$^4$($^3$P$^e$)3d is reduced, but this occurs not in a smooth fashion.
Instead, the figure suggests that the initial step in the ionization process occurs through a
coupling between 2s$^2$2p$^4$($^3$P$^e$)3d and 2s$^2$2p$^4$($^3$P$^e$)4p. Since the energy gap
between the states is about 1.7 eV, and hence comparable with the IR photon energy of 1.55 eV, this
coupling is expected to be strong. The ionization of the 2s$^2$2p$^4$($^3$P$^e$)3d may thus well
be governed by the ionization rate of the higher-lying 2s$^2$2p$^4$($^3$P$^e$)4p state.
It is important to note that the figure shows final-state populations, and that the figure
does not show the dynamics of the ionization process. The combined population of the two
states starts to reflect the oscillations in the ionization yield, but the modulation at a delay
time of 12 fs is too strong, and the modulations for delay times shorter than 8 fs are too weak.

Figure \ref{fig:3d} suggests that the atomic structure plays a significant role in the ionization
dynamics. The 2s$^2$2p$^4$($^3$P$^e$)3d and 2s$^2$2p$^4$($^3$P$^e$)4p states are strongly coupled,
and population may be transferred back and forth between these states. This suggests that, for
this peak IR intensity, a tunnelling picture may not give a complete description of the ionization
process for low-lying states. The tunnelling picture may be appropriate for higher states in the
Rydberg series, but the ionization of the 2s$^2$2p$^4$($^3$P$^e$)3d state appears to be
enhanced resonantly by the 2s$^2$2p$^4$($^3$P$^e$)4p state.

\begin{figure}
\includegraphics[width=7.9cm]{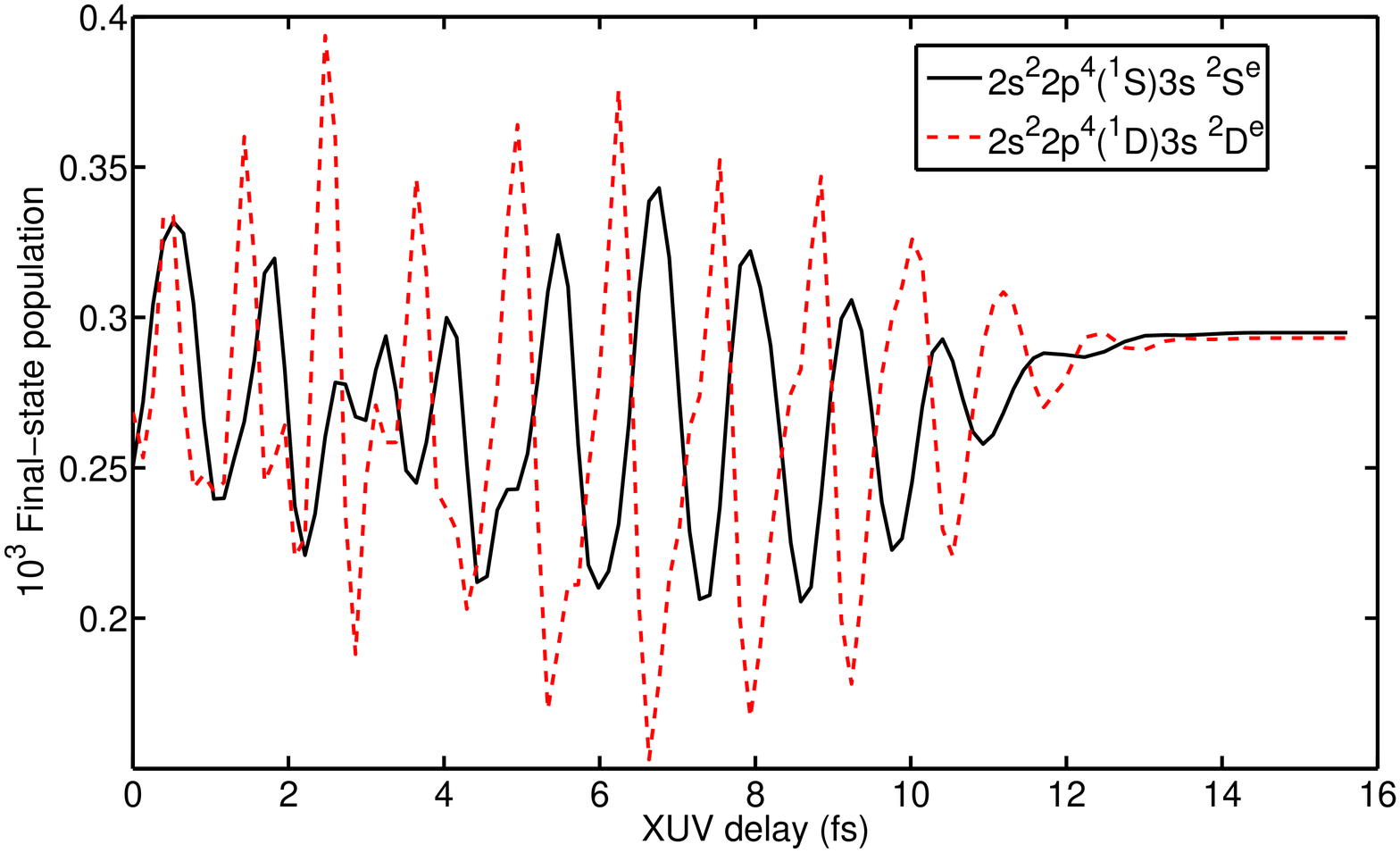}
\caption{(Colour online) Final-state population in 2s$^2$2p$^4$($^1$D$^e$)3s $^2$D$^e$
(red dashed line) and 2s$^2$2p$^4$($^1$S$^e$)3s $^2$S$^e$ (black solid line)
as a function of XUV time delay.}
\label{fig:3s}
\end{figure}

The most strongly bound electron to consider in detail is the 3s
electron. The final-state populations for the 2s$^2$2p$^4$($^1$S$^e$)3s $^2$S$^e$ and
the 2s$^2$2p$^4$($^1$D$^e$)3s $^2$D$^e$ states are shown in figure \ref{fig:3s}. These
populations remain within the range of 0.15 - 0.4$\times$10$^{-3}$ for all delay times. The
3s electron therefore
appears too strongly bound in either state to be ionized directly by the IR laser field. However,
the strong fluctuations in the final yield indicate that there is a strong coupling between these
states and other states. 
The binding energy of the 
2s$^2$2p$^4$($^1$S$^e$)3s is comparable to the binding energy of the 2s$^2$2p$^4$($^3$P$^e$)3d
state, but the former does not ionize whereas the latter does. It is thus not only the total
energy that is important, but also the threshold to which the electron is bound. 

Figure \ref{fig:3s} shows that the relative phase between the oscillations in the final yields for
the 2s$^2$2p$^4$($^1$S$^e$)3s $^2$S$^e$ and the 2s$^2$2p$^4$($^1$D$^e$)3s $^2$D$^e$ states changes
over time delay. At the longest delay times, the oscillations appear to be in anti-phase, whereas
for short delays, the oscillations are in phase. The origin of these oscillations is not obvious
to establish. Oscillations in anti-phase can be explained through population transfer between these
states themselves. The two states are linear combinations of the same uncoupled basis states,
involving a 2p$^4$ core with $m_{2p} = \{-1, -1, 1, 1\}$ and $m_{2p} = \{-1, 0, 0, 1\}$. A slight phase
change to one component can cause oscillations in the populations of 2s$^2$2p$^4$($^1$S$^e$)3s $^2$S$^e$ and
2s$^2$2p$^4$($^1$D$^e$)3s $^2$D$^e$ state. At shorter time delays, the population of the two states
is in phase. This suggests that these oscillations may be due to interactions with higher members of the
Rydberg series associated with the $^1$S$^e$ and $^1$D$^e$ threshold, as this would affect both 3s states
in a similar manner. Figure \ref{fig:3s} also suggests that for time delays of 2.5 - 5 fs, dynamics involving
the 2s$^2$2p$^4$($^1$S$^e$)3s $^2$S$^e$ may be especially interesting, as the oscillations are particularly
rapid.

\begin{figure}
\includegraphics[width=7.9cm]{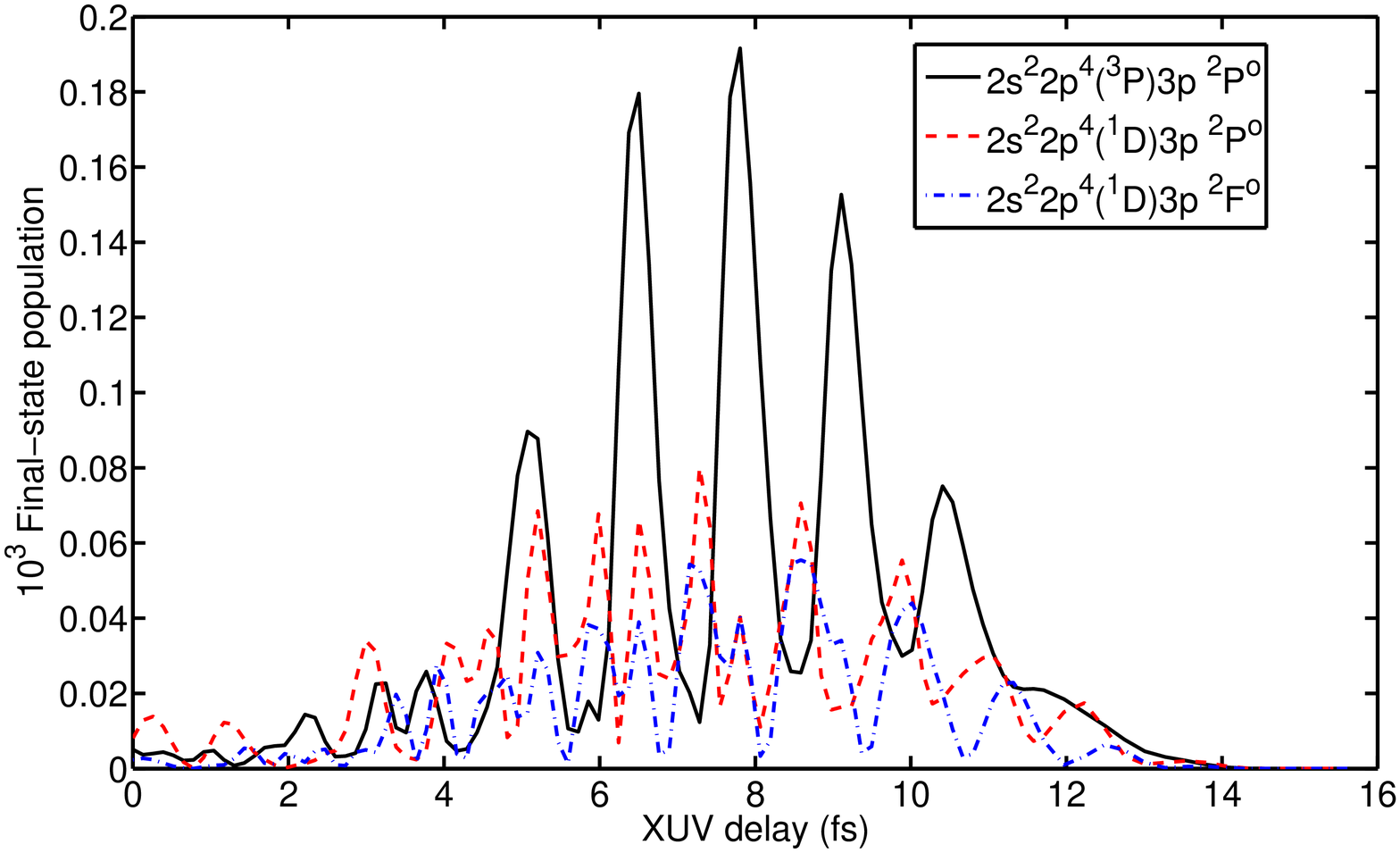}
\caption{(Colour online) Final-state population in 2s$^2$2p$^4$($^3$P$^e$)3p $^2$P$^o$ (solid black line),
2s$^2$2p$^4$($^1$D$^e$)3p
$^2$P$^o$ (red dashed line) and
2s$^2$2p$^4$($^1$D$^e$)3p $^2$F$^o$ (blue dot-dashed line)
as a function of XUV time delay.}
\label{fig:3p}
\end{figure}

Whereas excitation of 2p electrons to the 3s and 3d states by a single photon is possible, the
excitation of the 2p electrons to the 3p states by single-photon absorption is not allowed.
Nevertheless, due to the coupling between Rydberg states induced by the IR field, it is possible
for the 3p states to become populated during the time propagation. Figure \ref{fig:3p} shows
the final-state populations for these states. These
final-state populations tend to be strongest when the XUV delay time is near the peak of the IR
field. This suggests that these states are not being tunnel-ionized by the IR field. Instead they
need the IR field for population to be transferred into them. The mechanism here may be that the
2s$^2$2p$^4$($^3$P$^e$)3p $^2$P$^o$ receives an admixture of 2s$^2$2p$^4$($^3$P$^e$)3d
$^2$D$^e$ due to the IR field. The XUV field excites this mixed state. The IR field is insufficiently
strong to ionize this state, and when the IR pulse has finished some residual population is left
behind in the 2s$^2$2p$^4$($^3$P$^e$)3p $^2$P$^o$ state.

If all the populations in these strongly bound Rydberg states are taken into account, it becomes
possible to connect the modulations in the ionization yields with final-state populations in Rydberg
states. At the shortest time delays, the modulations are associated with population dynamics within
the 3s states. Near the peak of the IR field, low-lying 3p states can obtain an admixture of 3d
states. The excitation of the 3d states can then lead to a noticeable final-state population in the
3p orbitals, removing part of the excitation spectrum out of ionization pathways. When the XUV delay
increases further, the modulations become affected by interactions between states during the
ionization process. Rydberg state can couple strongly, leading to population not only being driven
towards higher-lying states, but also back towards lower-lying states. As a consequence, ionization
may not decrease uniformly with increasing XUV time delay.

\section{Conclusion}

In conclusion, we have used R-matrix theory including time dependence (RMT) to investigate IR-assisted
photoionization of Ne$^+$ using ultra-short pulses. Total ionization yields as a function of time
delay between the two pulses have been obtained. These yields demonstrate the fundamental nature
of the ionization process as excitation by the XUV pulse, followed by removal of the excited
electron by the IR pulse. The ionization yields do not increase smoothly: there are noticeable
oscillations on top of the ionization yield. For a full understanding of these ionization yields, a
multi-channel calculation is necessary: inclusion of just a single 2p$^4$ threshold does not provide the
full ionization yield.

Since the excitation of the ground state is independent of time delay, the oscillations reflect
dynamics occurring between Rydberg series and differences in the populations of excited final Ne$^+$ states.
We have demonstrated that there are significant variations in the populations of the residual Ne$^+$
states. The states responsible for the oscillations vary with time delay of the XUV pulse. When the
XUV pulse appears early in the IR pulse, the oscillations are primarily due to 3s electrons attached
to the Ne$^+$ thresholds, 3p electrons play a role when the IR field is at its strongest, whereas
oscillations when the XUV pulse is at the tail end of the IR pulse are associated with the emission
of, for example, 3d electrons due to the IR field.

The RMT approach has proven to be a suitable approach for the investigation of detailed atomic
structure effects on strong-field physics. Its foundation in R-matrix theory enabled us to obtain
a detailed atomic structure description of the Ne$^+$ ion, providing a detailed description of the
Rydberg series and the interplay between the different series. This detailed description led to
calculations where the inner-region description used a factor 10 more processors than the outer region.
The flexibility of the RMT approach allows us to systematically improve the description of the atomic
structure in the inner region, while keeping the number of channels in the outer region to a minimum.

The authors wish to thank M.A. Lysaght and L.R. Moore for their efforts in the development of
the RMT codes, and J.S. Parker for valuable discussions. This research
was sponsored by the Engineering and Physical Sciences Research Council (UK) under grant ref. no. G/055416/1
and through the EU Initial Training Network CORINF.
This work used the ARCHER UK National Supercomputing Service (http://www.archer.ac.uk).

\end{document}